\documentclass[lettersize,journal]{IEEEtran}
\usepackage{amsmath,amssymb,amsfonts}
\usepackage{algorithmicx}
\usepackage{algpseudocode}
\usepackage{algorithm}
\usepackage{array}
\usepackage{booktabs}
\usepackage[caption=false,font=normalsize,labelfont=sf,textfont=sf]{subfig}
\usepackage{textcomp}
\usepackage{makecell}
\usepackage{subcaption}
\usepackage{stfloats}
\usepackage{url}
\usepackage{verbatim}
\usepackage{placeins}
\usepackage{graphicx}
\usepackage{cite}
\usepackage{svg}
\usepackage{xcolor}
\usepackage{bbding}
\begin{document}

\title{TRANSOM: An Efficient Fault-Tolerant System for Training LLMs}

\author{\IEEEauthorblockN{Baodong Wu\IEEEauthorrefmark{1}\thanks{ Lei Xia and Qingping Li are the corresponding authors.}, Lei Xia\IEEEauthorrefmark{1}, Qingping Li\IEEEauthorrefmark{1}, Kangyu Li \IEEEauthorrefmark{1}, Xu Chen\IEEEauthorrefmark{1}, Yongqiang Guo\IEEEauthorrefmark{2}, Tieyao Xiang\IEEEauthorrefmark{1}, Yuheng Chen\IEEEauthorrefmark{1}, Shigang Li\IEEEauthorrefmark{3}} \\
\IEEEauthorblockA{
\IEEEauthorrefmark{1}SenseTime, \IEEEauthorrefmark{2}Huazhong University of Science and Technology, \\ \IEEEauthorrefmark{3}Beijing University of Posts and Telecommunications\\
Email: \IEEEauthorrefmark{1}\{liqingping, likangyu, chenxu1, xiangtieyao, yuhengchen\}@sensetime.com, \\
wubd.cs@gmail.com, xialei.thu@gmail.com, BaldStrong@hust.edu.cn, shigangli.cs@gmail.com
}
}



\maketitle
\begin{abstract}
Large language models (LLMs) with hundreds of billions or trillions of parameters, represented by chatGPT, have achieved profound impact on various fields. However, training LLMs with super-large-scale parameters requires large high-performance GPU clusters and long training periods lasting for months. Due to the inevitable hardware and software failures in large-scale clusters, maintaining uninterrupted and long-duration training is extremely challenging. As a result, A substantial amount of training time is devoted to  task checkpoint saving and loading, task rescheduling and restart, and task manual anomaly checks, which greatly harms the overall training efficiency.

To address these issues, we propose TRANSOM, a novel fault-tolerant LLM training system. In this work, we design three key subsystems: the training pipeline automatic fault tolerance and recovery mechanism named Transom Operator and Launcher (TOL), the training task multi-dimensional metric automatic anomaly detection system named Transom Eagle Eye (TEE), and the training checkpoint asynchronous access automatic fault tolerance and recovery technology named Transom Checkpoint Engine (TCE). Here, TOL manages the lifecycle of training tasks, while TEE is responsible for task monitoring and anomaly reporting. TEE detects training anomalies and reports them to TOL, who automatically enters the fault tolerance strategy to eliminate abnormal nodes and restart the training task. And the asynchronous checkpoint saving and loading functionality provided by TCE greatly shorten the fault tolerance overhead.
The experimental results indicate that TRANSOM significantly enhances the efficiency of large-scale LLM training on clusters. Specifically, the pre-training time for GPT3-175B has been reduced by 28\%, while checkpoint saving and loading performance have improved by a factor of 20.

\end{abstract}


\section{Introduction}



Undoubtedly, LLMs are currently the most trending topic in the field of AI. The use in various industries have propelled generative AI to unprecedented heights. Among them, the ChatGPT \cite{chatgpt} model developed by OpenAI stands as a typical example. It is already capable of assisting people in tasks like writing articles, generating code, and analyzing materials, fulfilling common text-language interaction needs. Its upgraded version, GPT-4\cite{gpt-4}, showcases even more astonishing reasoning capabilities. Following this, an increasing number of LLMs have been released, including Stanford Alpaca\cite{taori2023stanford}, Baidu's ERNIE Bot, Tsinghua's ChatGLM\cite{zeng2022glm130b}, Google's Bard, SenseTime's SenseChat, InternLM\cite{2023internlm}, Meta's Llama-1\cite{touvron2023llama}, Llama-2\cite{touvron2023llama2}, and more.


The success of LLMs can be attributed to several key factors. Firstly, the utilization of the transformer model \cite{xu2021anomaly} with self-attention mechanisms, which enhances the speed of model training, allows for the processing of longer sequences of data. Almost all LLMs employ the transformer as their neural network architecture. Secondly, the scaling of model parameters to the order of billions and the availability of high-quality annotated token datasets have played a pivotal role. According to OpenAI's research\cite{kaplan2020scaling}, LLMs exhibit the property of the ``scaling law" whereby their performance improves as training data and model parameter sizes increase, often leading to a sudden jump in capabilities at a certain scale. Lastly, the development of various hybrid distributed parallelization techniques has reduced the dependency of billion-parameter LLMs on the scale of GPUs. The Megatron-LM\cite{narayanan2021efficient} architecture proposed by NVIDIA optimizes transformer models for model parallelism, boosting training throughput by over 10\%. Microsoft's DeepSpeed\cite{rasley2020deepspeed} deep learning parallel optimization library accelerates training through various techniques like model parallelization, gradient accumulation, memory optimization, and mixed precision. It achieved training of a 13-billion-parameter LLM on a single NVIDIA V100 GPU and linear scalability on larger computing clusters.


As a result, larger parameter scales and more efficient training methods have become significant metrics claimed by model providers, computing platform suppliers, and chip manufacturers. For example, the 176B Bloom model\cite{scao2022bloom} completed training on a 350B dataset in 3.5 months using 384 A100-80GB chips; GPT-3 boasts a parameter scale of 175B, requiring 3640 PF-days of computation for training on a 300B dataset. DeepMind's Gopher\cite{rae2021scaling} model reached a parameter scale of 280B, while the multimodal PaLM\cite{chowdhery2022palm} model reached an astonishing 540B parameter scale. With the development and popularity of multimodal models, it's evident that training large models requires more computational resources for stable and continuous training. Model providers opt for self-built or leased high-performance heterogeneous AI computing platforms for large-scale pre-training or fine-tuning. NVIDIA GPUs, in particular, are favored in the field of large model training. Azure has built a cluster of over 10,000 GPUs for OpenAI, Cerabras has constructed an AI computing cluster named Andromeda\cite{Cerabras} with more than 2,512 H100 heterogeneous chips. SenseTime has built the SenseCore computing platform at the SenseTime Artificial Intelligence Computing Center as its computational foundation. This platform encompasses 27,000 GPUs, providing an impressive computational power output of 5,000 Petaflops for AI applications. According to the study by Sevilla et al\cite{sevilla2022compute}, the computational demand for large-scale AI models doubles approximately every 10 months. In the foreseeable future, the pursuit of more powerful hardware is expected to continue.


However, the expansion of computational scale introduces significant risks to the reliability of computing systems, particularly in the context of LLMs applications with demanding computational, storage, and communication requirements. This complexity poses challenges for achieving large-scale and efficient LLMs pre-training. We have summarized these challenges and attempted to address them through innovative approaches.


\textbf{Frequent anomalies in LLMs training tasks due to hardware and software issues.} These tasks often span hundreds of GPU nodes, and at such a scale, they are prone to various problems that can lead to task abnormalities within hours to days. For instance, according to the Bloom report, on a new cluster of around 400 GPUs, there are typically 1-2 GPU failures per week on average. Meta's 175B opt training records also demonstrate that within half a month, the 175B training experiment was interrupted more than 40 times due to hardware, infrastructure, and other issues. Additionally, even when a single node fails, processes on all other nodes of the task must be halted and the task needs to be killed and resubmitted. This process adds to the time overhead of the anomaly phase.


\textbf{Challenges in troubleshooting LLMs training tasks.} There are many reasons for anomalies in LLMs training tasks, such as node hardware failures, system malfunctions, network issues, storage problems, and training code errors, among others. We have compiled data on the causes of errors in some LLMs training tasks on the SenseCore cluster in Q2 of 2023, as shown in Table \ref{tab:errorstat}. Errors like ``NET/IB: Got completion from peer", ``socket timeouts", and ``GPU ECC errors" cannot be resolved solely through rescheduling; they require a deep analysis of which nodes are causing the errors and isolating the faulty nodes to resume training. Identifying the reasons behind timeout exceptions can be highly complex. For instance, the anomaly of communication timeout could be due to slow or faulty nodes, storage system malfunctions, or errors in the user's communication data sending and receiving code. Moreover, different reasons correspond to different recovery strategies, posing significant challenges for error diagnosis, pinpointing these issues often takes several hours or even longer.

\begin{table*}[t]
    \caption{
LLMs Training Task Error Statistics (From May 2023 to July 2023, running on SenseCore cluster)}
    \label{tab:errorstat}
    \centering
    \resizebox{0.8\linewidth}{!}{%
    \begin{tabular}{ccc}
    \toprule
    Error Categorization & Number of Tasks & Root Cause \\ \midrule
    Storage Read/Write Errors    & 34  & \makecell[l]{ Due to synchronization anomalies of storage servers, significant variations in the time overhead for file storage or object storage occur across different nodes. \\  This leads to communication waiting timeout or socket timeout in tasks. }  \\
    Network Communication Errors    & 43  & \makecell[l]{Incorrect insertion of IB network card; Uneven load distribution in RDMA traffic; \\ Misconfigured RoCE network switch; Expired ARP cache IP; Ethernet card or link is not connected.}  \\
    Node Hardware and Software Errors    & 66  & \makecell[l]{GPU ECC errors; GPU failure; Node not ready; Insufficient shared memory; Pod sends SIGTERM signal to exit; \\ Pod OOMKilled; Node image pull timeout; Local storage exceeding limit error.}   \\
    User Code and Training Environment Errors   & 179 &  \makecell[l]{Error in creating duplicate files with the same name; Data type conversion error; Python Segmentation fault; \\  CUDA runtime version mismatch with CUDA driver; AttributeError; torch.cuda.OutOfMemoryError; \\ RuntimeError; ModuleNotFoundError; NameError; AssertionError; OSError.}    \\
    Others   & 55  & \makecell[l]{System hang without error output; Occasional socket timeout errors without specific issues identified during troubleshooting; \\ Random occurrences of Pod processes with exit code -9 errors; Pod startup failures.}    \\ \bottomrule
    \end{tabular}%
    }
\end{table*}


\textbf{LLMs training tasks entail significant recovery overhead.} To mitigate the impact of the aforementioned anomalies on LLMs training, the current approach involves using checkpoints for recovery. The essence of checkpoints is to persistently store various data, including the optimizer state and weights, in the form of snapshots while the task is running. The total size of checkpoints is directly proportional to the parameter scale. For instance, a checkpoint of a 175B LLMs with FP32 optimizer state and bf16+fp32 weights amounts to 2.3TB. Training larger parameter-scale LLMs necessitates training on a large-scale GPU cluster, which requires increasing the frequency of checkpoint storage to minimize retraining time. For instance, in Bloom's 176B model training\cite{bloom176B}, a checkpoint is saved every 3 hours. In the case of the 175B OPT training\cite{opt175B}, a checkpoint is generated every 250 steps. However, as the model scale and the number of computational nodes increase, checkpoint read-write efficiency and stability become one of the main bottlenecks affecting training efficiency.


Based on the aforementioned challenges, we propose TRANSOM, a simple, efficient, and fault-tolerant large-scale model training system. The primary objective of TRANSOM is to provide an automated checkpoint-based fault-tolerant recovery pipeline system, significantly enhancing checkpoint access efficiency and fault recovery capability, reducing the cost of human intervention in troubleshooting training task anomalies, and improving the overall efficiency of large-scale model training tasks. In summary, this paper contributes in the following aspects:


\textbf{the Training pipeline Automatic Fault Tolerance and Recovery Mechanism:} TOL is a training pipeline automatic fault tolerance and recovery subsystem. It is based on a finite-state machine with a set of lifecycle management rules that enhance the management of distributed training processes within each working node during the training process pipeline execution. It introduces a distributed training process pipeline management mechanism in each worker node during the training process. Following the sequence of startup, warm-up, execution, verification, and recovery, this mechanism achieves dynamic and automated fault-tolerant recovery for training tasks. This system enables unattended closed-loop training, thereby enhancing the success rate of task startup.


\textbf{the Training Task Multi-dimensional Metric Automatic Anomaly Detection System:} TEE is a training task multi-dimensional metric automatic anomaly detection system. Starting from the moment the task is initiated, the training task's multi-dimensional metric automatic anomaly detection system continuously gathers relevant metrics. It employs a hybrid model and clustering algorithm to analyze the nodes and causes of training task failures. It then notifies the training pipeline's automatic management mechanism in the form of interruptions, triggering eviction and rescheduling of the faulty nodes. Simultaneously, the task reenters the initialization startup pipeline, continuing execution based on the latest checkpoint. This approach allows for precise identification and targeted handling of faulty nodes, effectively reducing the time and computational costs of manual troubleshooting and manual restarts.


\textbf{the Training Checkpoint Asynchronous Access Automatic Fault Tolerance and Recovery Technology:} TCE is a training checkpoint asynchronous access automatic fault tolerance and recovery technology. To achieve more efficient recovery of the training process, we propose a novel fault-tolerant asynchronous checkpoint access mechanism. This technology implements an asynchronous process from the initiation of checkpoint access operations in the training process to the persistence of checkpoints. It effectively avoids checkpoint loss due to single-point failures through redundant storage methods.

\section{Related Work}

\textbf{Training Large Language Models}. It is widely known that the model size of LLMs has been consistently increasing over time. For instance, the model size of GPT-3 \cite{brown2020language} in 2020 was 175 billion, and by 2022, PaLM \cite{chowdhery2022palm} had reached 540 billion. The substantial model size presents two challenges for training: first, GPU memory is limited and may not accommodate all parameters, necessitating strategies to run LLMs on clusters; second, the larger model size requires even larger datasets, prompting the need to accelerate LLM training.

Multiple parallelization approaches can meet and expedite LLM training, including data parallelism (DP), tensor parallelism (TP), pipeline parallelism (PP), and sequence parallelism (SP) \cite{li2021colossal}. DP involves having copies of the model on each device, with the dataset split across devices. Each device trains on its data and then contributes gradients for a global model parameter update. 

TP and PP fall under the category of model parallelism (MP), as they involve dividing the model across multiple devices due to its size. TP utilizes vertical splitting, while PP uses horizontal splitting.

SP is specifically designed for Transformers. Similar to DP, SP entails having identical model copies on each device, but it also falls under the umbrella of MP. A lengthy sequence is split into sub-sequences placed on different devices. These sub-sequences are then processed using Ring Self Attention, which leverages Ring AllReduce to synchronize Query and Key operations in self-attention.

Megatron-LM \cite{narayanan2021efficient} combines DP, TP, and PP into 3D parallelism, enabling the training of a trillion-parameter model on 3072 A100 GPUs in just 3 months. In contrast, ZeRO \cite{rajbhandari2020zero} achieves the training of a 100 billion-parameter model using only DP. ZeRO partitions parameter state, gradient, and parameter based on the dimension Nd of DP, with each DP rank storing only the relevant partitions, significantly reducing device memory overhead.

DeepSpeed \cite{rasley2020deepspeed} implements ZeRO and integrates it into 3D parallelism, coupled with mixed precision, ZeRO offloading \cite{ren2021zero}, and efficient checkpointing, making LLM training simpler and more efficient. Colossal-AI \cite{li2021colossal} optimizes TP, enhancing Megatron-LM's 1D TP to 2D, 2.5D, and 3D TP. Additionally, they introduce SP, enabling 4D parallelism. They also redesign the sharding and offloading module of DeepSpeed to achieve better performance. These LLMs training frameworks only focus on facilitating and high-performance execution of LLMs training, but they do not take into account fault tolerance and anomaly detection capabilities.







\textbf{Fault Tolerance for Large Language Models}. In large-scale LLMs training, encountering failures that lead to the collapse of the entire training process is a common issue, necessitating the use of fault tolerance mechanisms to address these problems. This area of work generally includes elastic training, elastic scheduling, and system fault tolerance.

In the realm of elastic training, MaTEx-Caffe \cite{amatya2017does} implements a plugin in Caffe that uses the MPIX\_Comm\_shrink method from the OpenMPI library to create a new communication group from the processes that did not fail when MPI\_Allreduce encounters a failure during gradient synchronization, achieving fault tolerance. Horovod \cite{sergeev2018horovod} Elastic manages all training processes through a centralized background Driver process, pausing and restarting all training processes by throwing and catching exceptions inserted into frameworks like PyTorch and TensorFlow. It also implements a memory checkpoint mechanism to ensure quick recovery after a restart. PyTorch \cite{paszke2019pytorch} Elastic starts an agent for each training process to manage it and achieves consensus on the number of nodes through a distributed consensus component. When one agent encounters an error, other agents retrieve the information from the consensus component and relaunch training processes based on the remaining available agents. Elan \cite{xie2020elan} has proposed a method for elastic communication based on the underlying device topology to achieve the scaling of training tasks. This approach enables IO-free training state saving and recovery.

In the direction of elastic scheduling, some works like \cite{xiao2018gandiva,peng2018optimus,peng2021dl2} exploit resource changes to adapt training and maximize resource utilization. DeepSys \cite{li2020scheduling} predicts the training speed of distributed deep learning jobs (DDL) using a neural network model and schedules DDL jobs based on this speed model. Pollux \cite{qiao2021pollux} periodically adjusts the batch size and learning rate of each deep learning job based on the training status of each job and the cluster's resource utilization. It also reallocates cluster resources for each job to optimize the average job completion time. Lyra \cite{li2023lyra} uses capacity loading and elastic scaling to balance the load between training clusters and inference clusters.

In the domain of system fault tolerance, Kubeflow \cite{kubeflow} utilizes the capabilities of Horovod Elastic and PyTorch Elastic to achieve dynamic scaling in distributed training on Kubernetes. Kubeflow Pipelines employ control flow to restart training upon failure until predefined conditions, such as accuracy demands or epoch requirements, are met. Apache Airflow \cite{airflow} is a more general pipeline tool. Singularity \cite{shukla2022singularity} has implemented a device proxy to intercept and forward GPU calls, enabling the time-sharing of GPUs for training jobs without user intervention, while ensuring a consistent number of visible GPUs on the user side. Efficient in-memory checkpoints are used to control overhead during time-sharing. Additionally, Singularity has implemented a distributed barrier for all reduce operation to ensure that there are no on-the-fly messages during checkpointing.

The aforementioned efforts lack an understanding of the behavior of training tasks and often resort to indiscriminate restarts without errors checks and troubleshooting. Additionally, many of these approaches intrude into deep learning frameworks \cite{amatya2017does,paszke2019pytorch,qiao2021pollux}, which can degrade the user experience. In contrast, this paper possesses both anomaly detection and error-checking capabilities while maintaining a non-intrusive stance towards deep learning frameworks.




\textbf{Anomaly Detection}. Anomaly detection can be viewed as a classification problem under the context of imbalanced data, typically involving unlabeled training data and falling under the realm of unsupervised learning. Various machine learning algorithms can be employed for anomaly detection. For instance, LOF \cite{breunig2000lof}, LOCI \cite{papadimitriou2003loci}, and LoOP \cite{kriegel2009loop} primarily detect anomalies by assessing the local data density. Isolation Forest \cite{liu2008isolation} partitions the feature space by constructing isolation trees to quantify the number of splits needed to isolate a sample. Feature bagging \cite{lazarevic2005feature}, as an ensemble detection framework, enhances detection performance by combining multiple models.

Anomaly Transformer \cite{xu2021anomaly} introduces an Anomaly-Attention mechanism to compute the association discrepancy and utilizes a minimax strategy to enhance the distinguishability of this discrepancy between normal and abnormal points. LSTM-VAE \cite{lin2020anomaly} utilizes LSTM for capturing temporal patterns and incorporates a Variational AutoEncoder (VAE) to perform reconstruction. OmniAnomaly \cite{su2019robust} extends the LSTM-VAE model by introducing a normalizing flow and utilizing reconstruction probabilities for anomaly detection.





\textbf{Checkpoint Performance Optimization}. LLMs training, due to its requirement for weeks or even months of training on large-scale clusters \cite{narayanan2021efficient,brown2020language}, necessitates regular checkpointing to ensure that training progress is not lost. Checkpoint optimization mainly involves two aspects: optimizing the content of checkpoints and improving checkpoint performance at the system level.

In the direction of optimizing checkpoint content, CPR \cite{maeng2021understanding} allows normal nodes to continue training in the event of failures on some nodes, employing partial recovery to restart the faulty nodes. By prioritizing the preservation of crucial model updates, it alleviates accuracy degradation issues, thereby reducing the overhead of storing and loading checkpoints. Vijay et al., 2022 \cite{korthikanti2023reducing}, selectively memory checkpoint transformer layers that occupy moderate memory and have reasonable computation loads, reducing the size of checkpoint data.

In the realm of system-level optimization, DeepFreeze \cite{nicolae2020deepfreeze} enhances checkpoint performance through asynchronous checkpointing, distributing checkpoints across multiple workers, and employing multi-level storage strategies. Nebula \cite{nebula} significantly reduces checkpoint time from hours to seconds by asynchronously saving checkpoints. A noteworthy highlight of this work is the incorporation of a fault-tolerant backup mechanism. In case of failures, all training participants do not need to restart from the beginning, which addresses a significant challenge in this domain.




\textbf{AI Training Platform}. In the industry, there are various solutions for AI job training, including Alibaba's PAI \cite{wang2019characterizing}, Huawei's ModelArts \cite{hu2021optimal}, ByteDance's Volcano Engine Machine Learning Platform \cite{volcengine}, Tencent's TI-ONE Training Platform \cite{ti-one}, Microsoft's Azure Machine Learning \cite{azureml}, and more. However, the majority of these platforms mainly provide basic support for LLMs training without incorporating features like fault tolerance, anomaly detection, or optimizing checkpoint performance.

We have found that ModelArts has done some work in the areas of fault tolerance and anomaly detection, but their implementations are relatively simple. The fault tolerance in ModelArts operates at the job level, whereas TRANSOM operates at the process level, making TRANSOM more efficient. ModelArts' anomaly detection only considers IO and GPU utilization to detect stuck anomalies, and the detection results are not integrated with fault tolerance capabilities. In contrast, TRANSOM utilizes monitoring metrics and logs to not only detect stuck anomalies but also pinpoint the problematic nodes. These detection results are then used by TRANSOM's fault tolerance tools to isolate problematic nodes and recover the training process.

\begin{table}[t]
    \caption{Approaches for LLMs training optimization.}
    \label{tab:ai-training-platform}
    \centering
    \resizebox{0.95\columnwidth}{!}{%
    \begin{tabular}{cccc}
    \toprule
    Approach & Fault Tolerant & Anomaly Detection & Checkpoint Optimization \\ \midrule
    DeepSpeed & \XSolidBrush   & \XSolidBrush  & \checkmark    \\ 
    PyTorch Elastic & \checkmark   & \XSolidBrush  & \XSolidBrush   \\
    Horovod Elastic & \checkmark   & \XSolidBrush  & \checkmark   \\
    Singularity & \checkmark   & \XSolidBrush  & \checkmark    \\
    DeepFreeze & \XSolidBrush  & \XSolidBrush & \checkmark    \\
    PAI & \checkmark  & \XSolidBrush & \XSolidBrush    \\
    ModelArts & \checkmark   & \checkmark  & \XSolidBrush    \\ 
    Azure & \XSolidBrush  & \XSolidBrush & \checkmark    \\
    TRANSOM & \checkmark   & \checkmark  & \checkmark    \\ 
    \bottomrule
    \end{tabular}%
    }
\end{table}

\section{TRANSOM overview}
\begin{figure*}[ht]
  \centering
  \captionsetup{justification=centering}
  \includegraphics[width=1\textwidth]{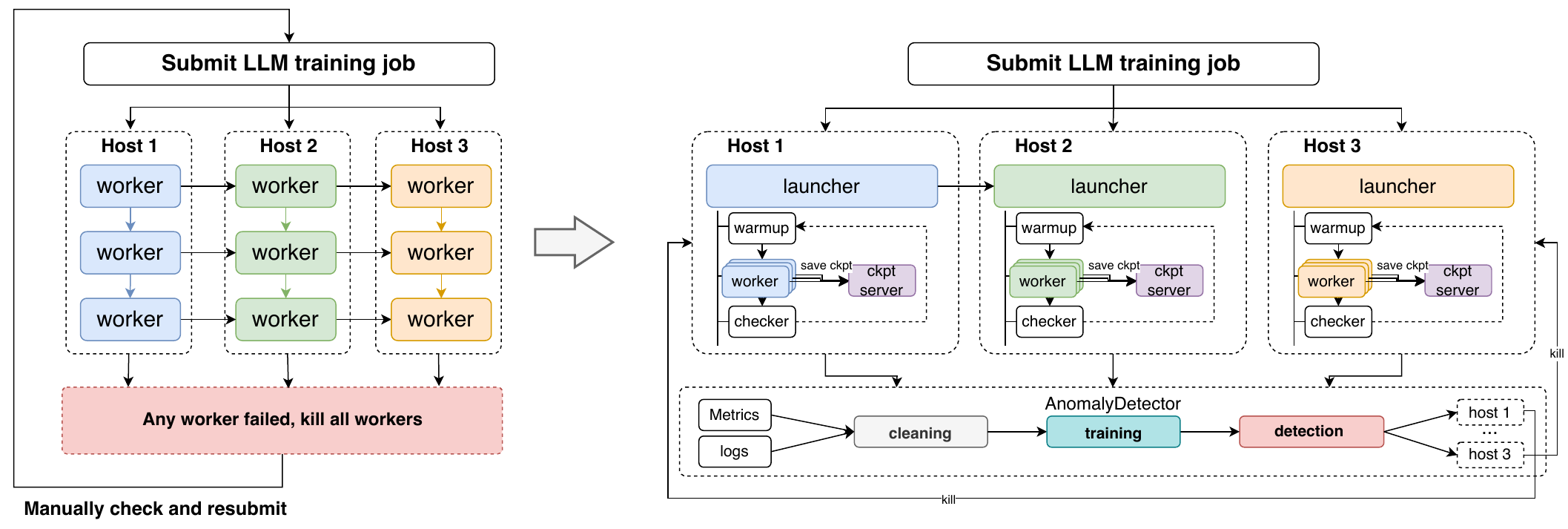} 
  \caption{TRANSOM Architecture} 
  \label{llm-acrt} 
\end{figure*}



In this section, we will introduce the overall architecture of the TRANSOM system, demonstrating how the proposed TOL, TEE and TCE subsystems work together.

As shown in the left half of Figure \ref{llm-acrt}, typically during LLMs training, if a failure occurs, all workers will be terminated, and manual restart of all workers is required. This approach has significant drawbacks: it incurs a high restart cost and can lead to the loss of considerable training progress if the checkpoint frequency is too low.



As illustrated in the right part of Figure \ref{llm-acrt}, we have conducted extensive optimizations to the entire LLMs training process. Firstly, considering the high frequency of errors during LLMs training, we have designed a mechanism to effectively manage the lifecycle of LLMs training jobs, named TOL. This is achieved through a sequence of steps: launch (step 1) -$>$ warm-up (step 2) -$>$ execution (step 3) -$>$ check (step 5, 6, 7) -$>$ recover (step 8 or step 9, 10, 11) to ensure the correctness of the training processes.

The most crucial aspect is the need for fault tolerance. Both PyTorch Elastic \cite{paszke2019pytorch} and Horovod Elastic \cite{sergeev2018horovod} provide fault tolerance and elasticity capabilities. However, given that changes in the number of workers participating in LLMs training can significantly impact convergence, the launcher is designed to have fault tolerance capabilities. During the training process, the anomaly detection service is continuously invoked to determine if any anomalies occur (step 5, 6), such as process blocking or node failures, etc. Upon detecting an anomaly, an error check is performed (step 7). If an anomaly node is detected, the training processes on the abnormal node are marked as failed. Subsequently, the launcher suspends all other training processes until the failed training processes are restarted. Then, the launcher restarts the entire training (step 9, 10, 11). If no anomaly nodes are detected, all training processes are restarted in place (step 8).


In step 5, when the anomaly detection service detects an anomaly in the training job, it notifies the launcher. The launcher then terminates all training processes and initiates an error-checking operation. In the anomaly detection service, we employ a novel hybrid anomaly detection approach, named TEE. This approach involves training a set of integrated anomaly detection models based on job logs and metrics data. These models are capable of detecting various anomalies, including freezing, communication issues, and code errors, among others. And report error information and related exceptional nodes to the launcher for quick recovery of the training.




In step 4, when the launcher terminates the training processes, it is necessary to checkpoint the model. This allows for a quick recovery from the point of failure when resuming the training. We have studied optimizations for reducing the checkpoint overhead at the system level. These optimizations primarily involve the use of memory caching mechanisms and asynchronous persistence mechanisms. The memory caching mechanism enables fast recovery from failed nodes by utilizing cached checkpoints. The asynchronous persistence mechanism allows for checkpoints without causing disruption to users and ensures the correctness of checkpoints, thereby accelerating the training efficiency of LLMs.


The detailed design aspects of the three aforementioned technologies have been documented in Section \ref{design}. For specifics on the automatic fault tolerance mechanism, please refer to Section \ref{tol}. Detailed information on the anomaly detection mechanism can be found in Section \ref{anomaly-detection}. The asynchronous checkpoint mechanism is elaborated upon in Section \ref{checkpoint}.
\section{TRANSOM design}\label{design}

\subsection{Training pipeline Automatic Fault Tolerance and Recovery Mechanism}\label{tol}


LLMs training typically requires a substantial amount of computation, networking and storage resources. The complete training takes several weeks or even months. For instance, the 176B BLOOM\cite{scao2022bloom} model takes 3.5 months to train on 384 A100 GPUs. In such large-scale scenarios, interruptions during training due to software or hardware failures are frequent. For example, an average of 1-2 GPU failures or issues like PyTorch deadlocks are encountered per week. These interruptions prolong LLMs training significantly. What is worse is that promptly detecting and resolving these problems on a distributed system by manual operations is challenging. As a result, resources are not efficiently utilized by training tasks.

\begin{figure}[ht] 
    \centering 
    \captionsetup{justification=centering}
    \includegraphics[width=0.5\textwidth]{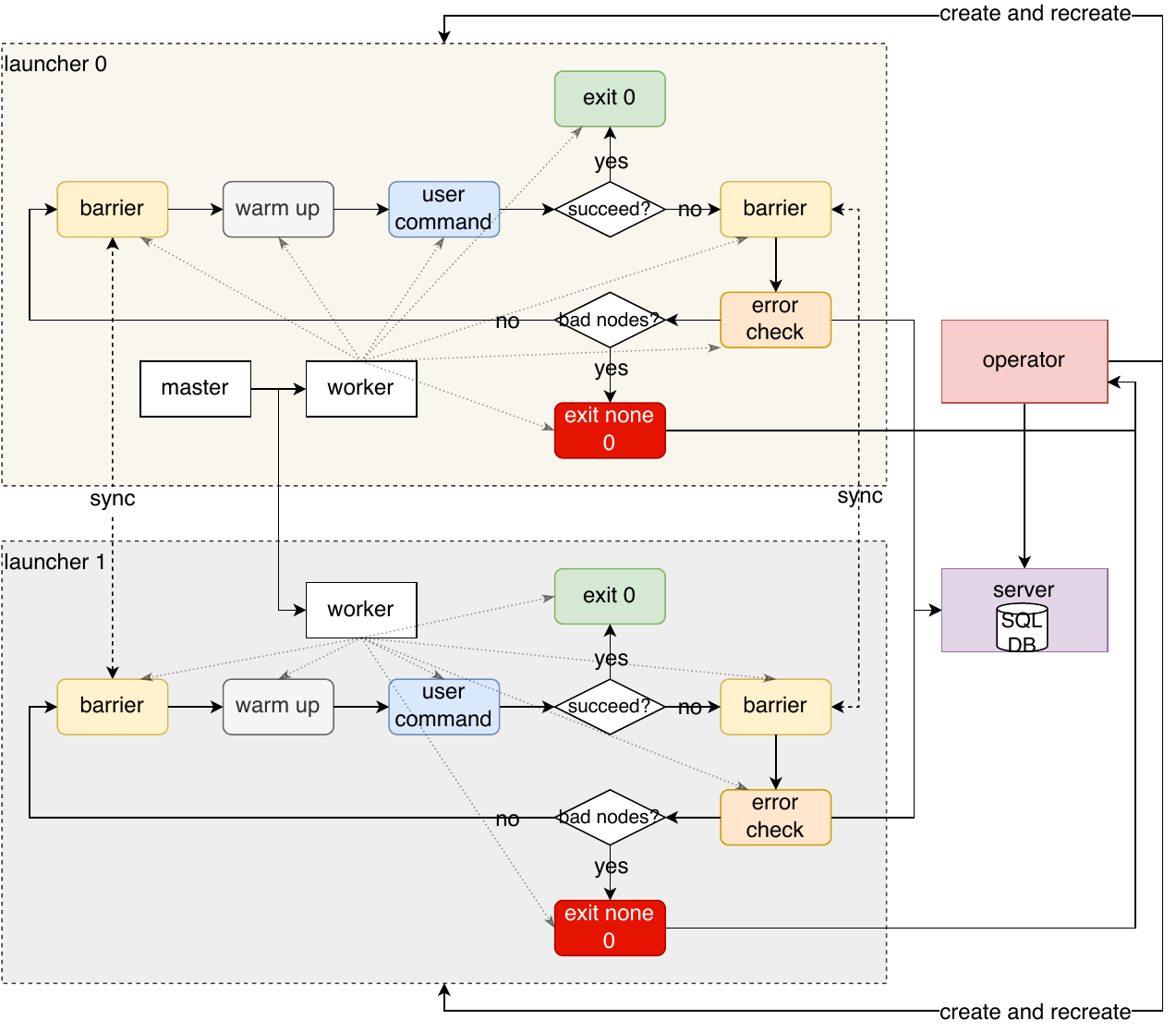}
    \caption{TOL modular composition} 
    \label{fig:launcher-arch} 
\end{figure}

We propose TOL to address this issue. TOL consists of a training task management tool called ``transom-launcher" and a fault-tolerant task orchestration controller named ``transom-operator". The launcher manages the lifecycle of training tasks. Upon encountering anomalies during the training, it temporarily suspends the training tasks, conducts error checks to identify the abnormal nodes, removes the problematic processes, and awaits the joining in of new training processes. The controller is responsible for creating launchers at the beginning of the training, and reconstructing new launchers after removing the problematic processes. In conclusion, TOL embodies the capabilities of error checking and fault recovery, significantly enhancing the efficiency of LLMs training. The architecture of TOL is illustrated in Figure \ref{fig:launcher-arch}.

From the architecture diagram of TOL, every training process is assigned a launcher for managing its lifecycle. The launcher process initiates before the initialization of training processes. The launcher further employs a leader election mechanism to distinguish between master and worker roles. On the master launcher, both the master and worker roles are activated concurrently, while on other launchers, only the worker role is activated.

The master role takes on the task of delegating assignments to the workers. These assignments encompass various tasks, including barrier synchronization, initiating training processes, conducting error checks, and halting training processes. Upon receiving a task, a worker promptly executes it and then reports back the outcomes. Further more, the master role, based on the results of tasks and user interactions, determines appropriate actions to distribute (such as notifying workers to exit, etc.). 


Next, we will introduce the tasks that the launcher master will distribute to each worker for execution.

\begin{enumerate}
    \item Infrastructure Warm-up Task: To ensure that the infrastructure is ready before the training processes start, we need to perform warm-up operations. These include disk checks (to ensure mounted datasets and code are prepared), GPU burn-in (to ensure each GPU device operates smoothly), connectivity checks (to ensure socket networking and InfiniBand networking between launchers is functional), and more. Infrastructure warm-up allows us to identify potential issues in advance, such as Socket Timeout due to network problems causing PyTorch errors.

    \item User Training Task: These are user-defined commands that need execution, such as scripts for training large models. This task takes precedence, with all other tasks serving to support its stable execution.

    \item Error Checking Task: When the user training task encounters errors, the error checking task swiftly pinpoints the source of the issue. Error checking operations involve disk checks (to verify accessibility of mounted datasets and code), GPU burn-in (to identify any malfunctioning GPUs), NCCL test (to verify GPU-to-GPU communication), and anomaly node detection (calling the anomaly detection service to identify problematic nodes, more details in section \ref{anomaly-detection}).
\end{enumerate}

Typically, these three types of tasks are executed in the following order: Infrastructure warm-up task, user training task, error checking task. During the execution of the user training task, the master periodically queries the anomaly detection service to check if the training task has encountered anomalies. Upon detection, the master distributes the error checking task. The master need to query the anomaly detection service regularly due to instances of the training task becoming stuck or encountering errors without exiting are scenarios where the master cannot directly detect anomalies. However, the anomaly detection service can identify these issues. Therefore, the master needs to retrieve the actual state of the training task from the anomaly detection service.



Through the above actions, the master identifies the abnormal nodes. If the abnormal nodes are empty, it implies that the interruption in the training tasks might have been due to software issues. In this case, the master will perform a local restart of all training tasks. However, if there are abnormal nodes, these nodes need to be excluded to avoid affecting the future training. At this point, the master will terminate the training processes on the abnormal nodes and report the abnormal nodes to the transom-server. Then, it waits for the creation of new launchers on fresh nodes.

The transom-operator is responsible for creating the new launchers and sets node anti-affinity policies based on the information obtained from the transom-server. This ensures that the newly created launchers are not scheduled to the abnormal nodes.

In summary, TOL equips LLMs training with fault tolerance and recovery capabilities. It leverages the anomaly detection service and error checking tasks to assist in identifying abnormal nodes, reducing the need for manual intervention and thus enhancing the efficiency of LLMs training.

\subsection{Training Task Multi-dimensional Metric Automatic Anomaly Detection System}\label{anomaly-detection}


During long and large-scale training processes, hardware and software failures are unavoidable. In such cases, constructing an automated, interpretable, and high-quality anomaly detection system would greatly assist in improving operational efficiency, thus improving training efficiency.


To address this, we introduce TEE, a multi-metric anomaly detection system designed to swiftly detect abnormal behaviors of training tasks and pinpoint the underlying causes of anomalies. TEE capitalizes on the prior features of training tasks by employing a variety of metrics in combination, encompassing both log-based and metric-based detection. The diagram \ref{fig:anomaly-detection-arch} below illustrates the components of TEE, which is divided into two subsystems: offline training subsystem and online detecting subsystem.

\begin{figure}[ht] 
    \centering 
    \captionsetup{justification=centering}
    \includegraphics[width=0.5\textwidth]{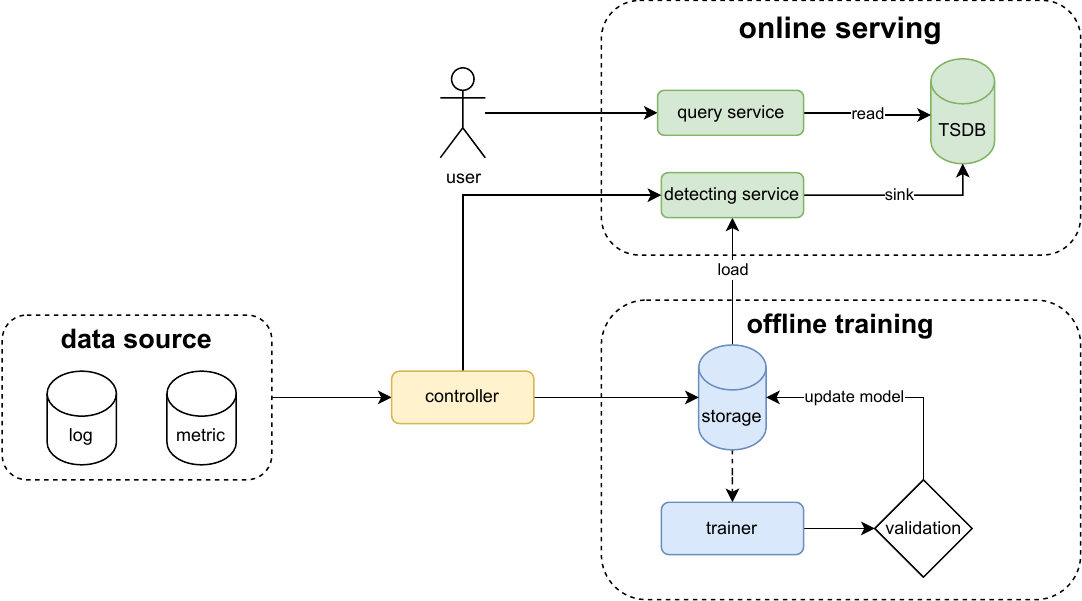}
    \caption{TEE modular composition, consisting of offline training subsystem and online detecting subsystem} 
    \label{fig:anomaly-detection-arch} 
\end{figure}

\begin{figure}[ht] 
    \centering 
    \captionsetup{justification=centering}
    \includegraphics[width=0.48\textwidth]{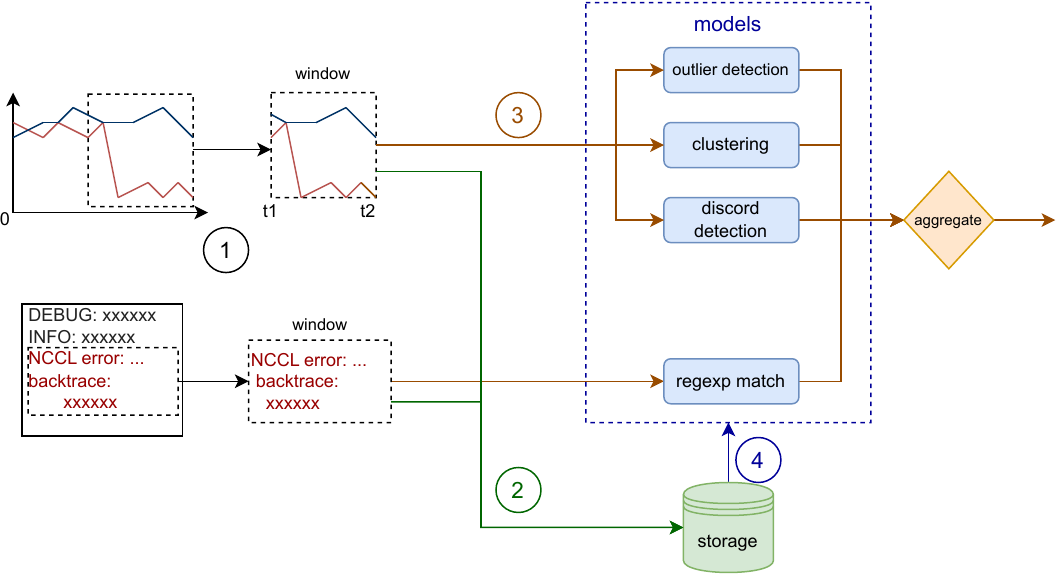}
    \caption{TEE system workflow: 1. source data capture; 2. offline dumping; 3. online detection for each window; 4. model iteration} 
    \label{fig:anomaly-detection-procedure} 
\end{figure}

The system workflow is roughly illustrated in the diagram \ref{fig:anomaly-detection-procedure}. Raw data from log system and monitoring system are captured. On one hand, they are dumped into the offline training subsystem, where they undergo preprocessing before being used for subsequent offline model training. On the other hand, they enter the online training subsystem, where multiple internal models process the data separately and aggregate the final detection results. Additionally, the models will be continuously updated to improve accuracy and performance.

\subsubsection{Offline Training Subsystem}


This subsystem is responsible for collecting log and metric data, then annotating, cleaning and preprocessing it before storing data into distributed storage for subsequent model training. 


In both the log-based and metric-based detection methods, we deliberately opted for more interpretable traditional machine learning and statistics methods. There were two primary reasons for this decision. Firstly, we encountered challenges related to data quality. For instance, during normal training, there is a minimal volume of logs, but when errors occur, a massive amount of logs is generated. Secondly, deep neural networks lack interpretability, making it difficult to pinpoint the root causes of anomalies. The detail of logging and monitoring detection is presented as follows.


\textbf{Log-based detection}: A sliding window is adopted to keep track of the number of error logs within each. When the number of error logs exceeds a certain threshold, the task is considered as anomalous. Also, based on the observation that in the event of a task error, the node that first produces error logs is often the actual anomalous node. We utilize the results of log analysis for locating the anomalous nodes.


\textbf{Metric-based detection}: We leveraged three key characteristics of LLMs training as prior knowledge, and based on each of them, an appropriate model is adopted.

\begin{enumerate}
    \item The behavior of various ranks is statistically consistent, which suggests the use of clustering models. For simplicity and low-latency, the straightforward K-nearest neighbors (KNN) \cite{fix1989discriminatory} method is adopted. To perform clustering on time series data, we employed Dynamic Time Warping (DTW) \cite{gold2018dynamic} technology.
    \item Individual ranks exhibit periodic behavior, which indicates the need for algorithms capable of detecting periodicity. We initially attempted to use matrix profile \cite{yeh2016matrix}, which is capable of identifying both periodicity and anomalies within a time series segment. But its performance was subpar due to its reliance on single nearest neighbor, which is inaccurate on detecting anomalies. So we adopted the KNN matrix profile algorithm \cite{he2020neighbor}. Compared to matrix profile, this approach significantly improved detection accuracy.
    \item Multidimensional metric at each timestamp exhibits classifiable characteristics, which suggests the need for an outlier detection algorithm. Common algorithms for this purpose include LOF (Local Outlier Factor) \cite{alghushairy2020review} and Isolation Forest \cite{liu2008isolation}. Between them, Isolation Forest is more suitable for detecting global anomalies, while LOF excels at identifying local anomalies. LLMs training can be divided into at least training and checkpoint stages, and considering the variability of different frameworks, algorithms, we opted for the LOF algorithm.
\end{enumerate}


TEE stores the trained model in a distributed storage and periodically engages in retraining. The dataset is divided into a training set and a testing set. The testing set is utilized to test the key metrics of the model, such as accuracy, precision, and recall. The model is evaluated by these metrics. Versions of the model that do not pass the testing are discarded.

\subsubsection{Online Detecting Subsystem}


Online detecting subsystem operates by loading the trained models from storage and providing an API for executing anomaly detection. The detection service is invoked periodically for each actively running task to analyze the logs and metric data generated during that specific time frame. During the detection, all models firstly operate independently, and the detection service aggregates their results to produce the final detection outcome. The pseudocode is shown in Algorithm~\ref{alg:anomaly-detection}.

\begin{algorithm} 
\caption{anomaly detection procedure pseudocode}
\label{alg:anomaly-detection}
\begin{algorithmic}[1]
    \Function{log\_detection}{$L_t$}
        \State $err\_cnt \gets 0$
        \While{$line \gets L_t$}
            \If{$isErrorLog(line)$}
                \State $err\_cnt += 1$
            \EndIf
        \EndWhile
        \State \Return $err\_cnt > thd$
    \EndFunction

    \Function{metric\_detection}{$M_t$}
        \State $r1 \gets LOF(M_t)$
        \State $r2 \gets NeighborProfile(M_t)$
        \State $r3 \gets Clustering(M_t)$
        \State \Return \Call{aggregate}{$r1$,$r2$,$r3$} 
    \EndFunction
    
    \State $M \gets multi\_dimension\_metrics$
    \State $L \gets stream\_logs$
    \State $W \gets window\_size$
    \While{$t \gets time.now()$}
        \State $M_t \gets M[t-W,t]$
        \State $L_t \gets L[t-W,t]$
        \State $r1 \gets$ \Call{log\_detection}{$L_t$}
        \State $r2 \gets$ \Call{metric\_detection}{$M_t$}
        \State \Return $r1 | r2$
    \EndWhile
\end{algorithmic} 
\end{algorithm}


When the system detects anomalies, it attempts to collaborate with the log system to identify error nodes. However, it's important to acknowledge that the current capabilities of TEE in diagnosing error root causes are limited. It can only indicate anomalous nodes or ranks to users and cannot provide in-depth information. The confirmation of error nodes also relies on the TOL system mentioned in the previous section \ref{tol}.

\subsection{Training Checkpoint Asynchronous Access Automatic Fault Tolerance and Recovery Technology} \label{checkpoint}


As mentioned previously, it is necessary to decrease the latency of saving and loading checkpoints. When saving or loading checkpoints, the performance is limited by the network attached storage, which is highly inefficient. To address this issue, We propose TCE, a lightweight design to reliably and efficiently save and load checkpoints.


\subsubsection{TCE overview}

\begin{figure}[ht] 
    \centering 
    \captionsetup{justification=centering}
    \includegraphics[width=0.5\textwidth]{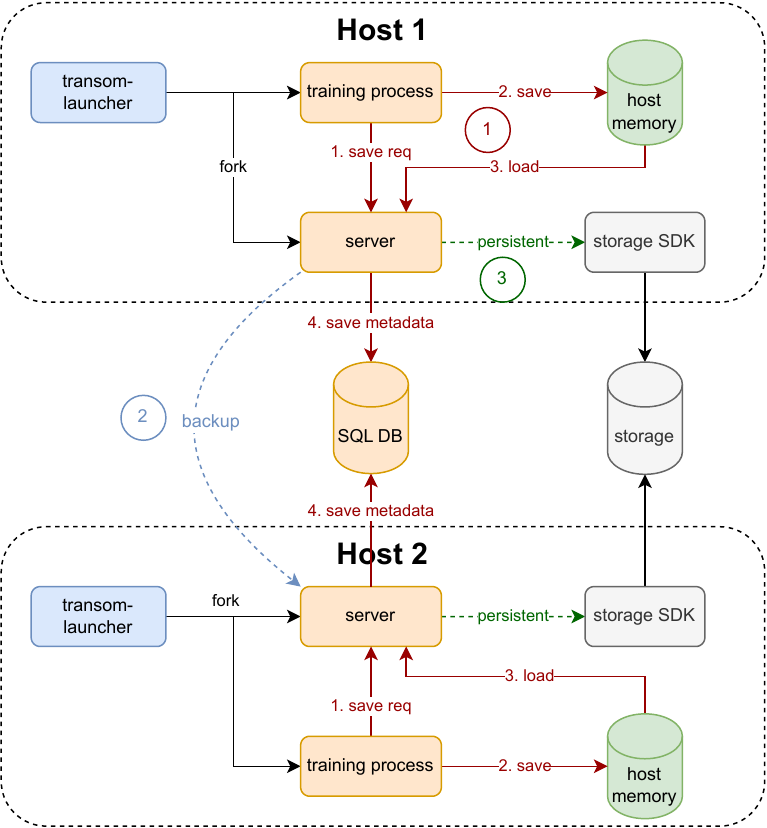}
    \caption{TCE modular composition and workflow overview} 
    \label{fig:ckpt-arch} 
\end{figure}

It primarily leverages an in-memory caching mechanism and asynchronous persistence, supplemented by cache eviction and cache backup. The figure \ref{fig:ckpt-arch} below describes the modular composition and general workflow of TCE. It forwards checkpoint save and load request to server, enabling completely control over checkpoint data. A server is adopted to handle these requests. By caching checkpoint data in memory, performance is magically improved. The key features of TCE is presented as follows.

\textbf{Ultra-low latency}: TCE introduces an in-memory cache to temporarily hold checkpoint data, avoiding inefficient disk I/O and network transmission during checkpointing. Checkpoints are first written to the cache and persisted asynchronously to reliable storage, hiding the latency of slow persistence operations. By optimizing the checkpoint lifecycle in this manner, TCE can significantly reduce checkpoint latency, thus improve training efficiency.

\textbf{Excellent fault tolerance}: The volatile nature of in-memory caching poses a challenge for fault tolerance in our approach, especially when failed containers are rescheduled and lose their in-memory caches. To address this issue, we leverage RDMA high-speed networking for cache backup. After scheduling all pods, TCE maps the training pod list to node ranks, with each TCE server on a given node rank asynchronously and durably backing up its checkpoint cache to the TCE server on the next node in sequence. TCE servers attempt to autonomously restore any lost caches after recovery by fetching the backup data from the previously backing-up node and notifying the former node to re-backup. Barring near-simultaneous failures of adjacent nodes, training processes can then rapidly recover from the memory caches. In reality, almost all infrastructure failures are single-node incidents, with multi-node damage only occurring at the rack, room or data center level - events with comparatively rare probability. Therefore, TCE can provide excellent fault tolerance for checkpoint caching systems.


\textbf{Flexibility on memory usage}: Given the finite nature of container memory, TCE also supports a combination of two cache eviction strategies. The first is that a memory usage limit can be specified. When memory is insufficient, TCE prioritizes evicting the oldest cache. The second is that users can specify the maximum number of checkpoint cycles that TCE can cache. Above strategies provide more flexibility and control over memory usage.

\subsubsection{Theoretical Performance Analysis}


In this section, we will demonstrate the superiority of our solution through formula derivations.


Let us take the example of a 175B LLMs trained with the Adam optimizer, employing ZeRO optimization, and using FP16 for training. Assuming parameter count is denoted as $P$, training with $N$ nodes and each node has 8 GPUs. In terms of parallelization, denote $TP\_SIZE$ as $TP$, $DP\_SIZE$ as $DP$, $PP\_SIZE$ as $PP$, and they satisfy the equation $DP \times PP \times TP = 8N$. Additionally, for ease of calculation, assuming that the optimizer and model data size on each rank is consistent and if $DP$ is greater than 8, it is always a multiple of 8.


\textbf{Saving checkpoint}: The model's weights require $2 \times P$ bytes of storage, in total $\frac{8 \times N}{DP}$ ranks save them. The optimizer's state requires $12 \times P$ bytes of storage. Denote data size saved by each rank as $S_{rank\_save}$, it can be deduced that the maximum amount of data saved by a single rank is

\begin{equation}
\begin{aligned}
    max(S_{save})=\frac{2  P}{\frac{8  N}{DP}}+\frac{12  P}{8  N} \\
    =\frac{(DP+6)  P}{4  N}
\end{aligned}
\end{equation}


In both the traditional approach and TCE, $max(S_{save})$ is the same. So the performance gain, denote as $G_{save}$ can be represented as

\begin{equation}
    G_{save}=\frac{B_{mem}}{B_{nas}}
\end{equation}

where $B_{nas}$ is the average write speed into network attached storage, and $B_{mem}$ is that into local memory. We'll analysis the bandwidth later.


\textbf{Loading checkpoint}: Each rank not only reads the optimizer state written previously but also needs to read the model state, resulting in larger load size than written size. Load size of Each rank is the same as $max(S_{save})$.


In the traditional approach, reading additional data leads to increased latency. However, TCE can accelerate reading through RDMA communication (in the case that model state is cached by another node) and deduplicate redundant requests (when 8 ranks request the same model state from another node, there will be only one flying request), significantly improving reading performance. The latency of loading by TCE is deduced to

\begin{equation}
\begin{aligned}
    T_{load}=\left\{ \begin{aligned} & \frac{(DP+6)  P}{4  N \times B_{mem}} && DP <= 8  \\ &\frac{3  P}{2  N \times B_{mem}} + \frac{(DP-8)  DP \times P}{32  N \times B_{rdma}} && DP>8 \end{aligned} \right.
\end{aligned}
\end{equation}

where $B_{rdma}$ is the RDMA network bandwidth of each node.


\textbf{Performance analysis}: It's important to note that TCE performs much better than traditional approach. The limit of network attached storage is as follows:

\begin{itemize}
    \item Performance of network attached storage is much worse than its of memory. Network attached storage usually uses ethernet, whose bandwidth is much smaller than memory bandwidth. For example, typical ethernet bandwidth is 10Gbps, while total bandwidth of DDR4 memory is about 20GBps.
    \item In most cases, performances of each rank are not consistent, resulting to slow ranks harming the overall latency. For example, we have noticed that some training tasks get stuck due to the latency of saving checkpoint exceeds PyTorch barrier timeout, which is 30 minutes.
    \item The performance is limited as node scale grows. It inevitably leads to bandwidth race at server side.
\end{itemize}

TCE addresses all above issues. Firstly, it utilizes in-memory cache, which minimizes user-aware latency. Secondly, memory bandwidth is so enough that even 8 ranks save checkpoints simultaneously, the bandwidth is not used up. As a result, even the slowest rank achieves high performance. The last but not the least, TCE is magically scalable by utilizing local memory and lossless fabric. When the node scale increases, the performance of each rank is not affected. TCE even benefits from large-scale training task, due to the reduction in the amount of data read and written per rank.

In real world scenarios, saving a general checkpoint of a 175B model with 128 ranks and DP set to 8, takes approximately 4.5 minutes on SenseCore's file storage, roughly estimating an average write bandwidth of around 71.1MB/s per rank. In contrast, TCE can complete the same operation in 10 seconds for the same scale, achieving a 27-fold performance gain.

\section{Implementation}




All subsystems, namely TOL, TEE and TCE, are designed to function on Kubernetes. TOL is implemented in golang, providing fault tolerance for LLMs training. TEE is implemented in python, providing anomaly detection functionality. At last, TCE is implemented in C++, providing extreme high performance for checkpointing. The engineering highlights of each subsystem are explained in detail as follows.

\subsection{Highlights of TOL}


\textbf{Compatibility across OS distributions}: We've developed all the components of the TOL subsystem Using Golang. Golang's inherent feature of static compilation ensures that the resulting binaries are highly self-contained, maintaining good compatibility across various operating system distributions.


\textbf{Debugging convenience}: We've indeed provided users with the capability to manage tasks effectively. Inside the task container, the launcher exposes an RPC service. When users issue commands, an RPC request is sent to the launcher, which handles the processing. This feature enables users to manually use the CLI to start or stop training job, and inspect the environment, significantly enhancing debugging convenience.


\textbf{Light-weighted leader election algorithm}: Given that tasks among multiple pods are not independent, launcher require mutual awareness and collaboration. We've implemented a specialized leader election algorithm for the launcher, independent of external database or kubernetes lease. 

We utilize a stateless transom-server. Launchers periodically send requests to the server to compete for leases. The server maintains a simple in-memory map containing all active leases. Transom-server can tolerate server downtime without interrupting training process. When a launcher initiates a lease request, it always carries the information from the previous lease, so that even after the service restarts, it automatically renews the previous lease.

\subsection{Highlights of TEE}


\textbf{Data pre-processing}: The key of training is data pre-processing. In summary, we solved three main challenges.

\begin{enumerate}
    \item Determine the value of metrics. The system involves a plethora of metrics related to training. Some data may be unrelated to training, while others might exhibit linear correlations. To address this challenge, a combination of domain knowledge and correlation analysis is used. By analysing previous training job metrics, relevant indicators are selected. Additionally, by conducting correlation analysis, potential strong linear relationships between different metrics are identified and utilized as well.
    
    \item Distinguish useless time series data. Before training actually begins, there are multiple steps involved, including container initialization, framework initialization, dataset loading, and more, whose metrics are meaningless. We utilize manual annotation to separate meaningful data from this initialization phase for training purposes.
    
    \item Certain metrics experience extremely rapid fluctuations that are not directly usable. Inter-node IB network traffic and intra-node nvlink traffic often undergo transitions between 0 and 1 after normalization. This phenomenon arises from the rapid pace of a single forward-backward pass during training, contrasted with the relative long interval at which metrics are collected. While these transitions may hint at periodicity, according to the Nyquist–Shannon sampling theorem \cite{1697831}, we cannot perfectly reconstruct these fluctuations without distortion. Our approach is to employ median filtering for signal smoothing.
\end{enumerate}


\textbf{Model implementation and uutcome aggregation}: Our model code is written in Python and utilizes open-source Python libraries such as sklearn, scipy, tslearn, etc. Multidimensional metric and multiple models are utilized for better effectiveness. Since we employ three models, each request yields three fundamental results, which need to be used to compute the final outcome. The current strategy is to consider a task as anomalous if at least two of the models classify it as such.


\textbf{Periodic and manual detection}: Training jobs are detected periodically, so that anomaly could be detected as early as possible. Meanwhile, manual detection can be triggered by end user. Detection results are considered as time-series data, storing in a popular TSDB named Prometheus.

\subsection{Highlights of TCE}


\textbf{Zero overhead to end-user}: The TCE client is implemented in Python and packaged as a Python package. Users can enable TCE functionality simply by running the command \verb|pip install transomSnapshot| and adding an \verb|import transomSnapshot| line to their code. Notably, to minimize user burden, we utilize monkey patch to adapt to DeepSpeed \cite{rasley2020deepspeed}. This means users don't need to manually modify the DeepSpeed framework and can maintain compatibility even with customized frameworks.



\textbf{Ultimately low latency}: In the implementation of TCE, the most critical aspect is achieving efficient IO. We utilize three key optimizations.

\begin{enumerate}
    \item multi-thread memory copy: single-thread memory copy is limited to around 2GBps. So it is urgent to use multiple threads for concurrent IO.
    \item DMA operations for intra-node communication: TCE employs relatively small pinned memory as the intermediate buffer for IPC. Data is firstly copied to buffer by chunk, then copied to destination. By utilizing DMA mechanism, memory IO is improved efficiently. The reason behind this improvement is the substantial reduction in CPU cache misses.
    \item RDMA for inter-node communication: By utilizing high performance RDMA network and support multiple NIC working simultaneously, inter-node loading is accelerated, ensuring consistently high performance.
\end{enumerate}

The memory copy optimization is worth explaining in detail. Denote $N$ as tensor size, $n$ as number of threads and $k$ as buffer size, the execution logic for a cuda tensor is listed in Algorithm \ref{alg:memcpy}.

\begin{algorithm} 
\caption{Optimized memory copy for cuda tensor}
\label{alg:memcpy}
\begin{algorithmic}[1]
    \For{$i \gets 1\ to\ n$}
        \State $buf_i \gets cudaHostAlloc$
    \EndFor
    \For{$i \gets 1\ to\ n$}
        \Comment{run following code in seprate thread}
        \State $beg \gets N/n*i$
        \State $end \gets N/n*(i+1)$
        \For{$j \gets beg/k\ to\ end/k$}
            \State $src \gets tensor+k*j $
            \State $dst \gets host\_addr+k*j$
            \State \Call{$cudaMemCpy$}{$buf_i$, $src$, $k$}
            \State \Call{$memcpy$}{$dst$, $buf_i$, $k$)}
        \EndFor
    \EndFor
\end{algorithmic} 
\end{algorithm}


\textbf{Linux memfd adoption}: Regarding the IPC mechanism between client and server, TCE uses the memfd mechanism in the Linux kernel, rather than traditional shared memory. There are three reasons for this. First, the shared memory capacity within containers is limited. Even with special configurations, shared memory will not exceed the maximum value set in the kernel (typically half of the physical memory). This limitation might prevent the full utilization of available memory. Second, the user's software stack has unrestricted access to shared memory, which could potentially lead to contention with TCE, compromising the user's training tasks. Third, to make better use of CPU memory bandwidth, we use the hugepage mechanism to reduce page faults. Memfd provides a more convenient way to use hugepages, whereas traditional shared memory requires settings in /dev/hugepages to use hugepages. TCE utilizes memfd, which uses pure heap memory, to share memory among multiple processes, avoiding the issues associated with shared memory as mentioned above.


\textbf{Final state consistency}: At last, to ensure the correctness of asynchronous backup and persistent logic, TCE draws inspiration from Kubernetes' principles of declarative management and has developed a C++ version of a Kubernetes operator. By simply adding the checkpoint file name to the task queue, the reconciliation logic ensures the correctness of the desired final state within the TCE server.

\section{Evaluation}


In this section, we conducted comparative experiments using TRANSOM on the Sensecore GPU cluster.



Our experiments yielded the following key results:

\begin{enumerate}
    \item The end-to-end training time for the GPT3-175B model on 512 A100 GPUs has been reduced from 118 days to 85 days, resulting in an overall training efficiency improvement of 28\%.
    \item The average time spent on restarting the task after an error has been reduced from hours to 12 minutes, resulting in effective training time accounting for over 90\% of the end-to-end training time..
    \item The checkpoint save and load latency for the GPT3-175B model has been reduced from an average of 200 seconds to less than 10 seconds.
    \item The multi-indicator anomaly detection method increased the success rate of detecting anomalies in large models to 70\%, achieving 100\% coverage of error types, with detection times in seconds.
\end{enumerate}


\subsection{Experimental Setup}



Our experiments were conducted using a combination of real-world training and mathematic simulations for large-scale scenarios. The training run on Sensecore AI cloud. Each computing node is equipped with a 112-core CPU, 1024 GB of DRAM, and 8 NVIDIA A800-80GB heterogeneous chips. Nodes were connected to each other by a high-speed 200Gbps InfiniBand network and one 25Gbps Ethernet network. Additionally, we constructed a simulator based on traces from a subset of LLMs training tasks as shown in \ref{tab:errorstat}. This simulator was used to compare the training efficiency and anomaly detection efficiency with and without the use of the TRANSOM system.

To train the GPT model, we adopted a combination of NVIDIA's Megatron-LM and Microsoft's DeepSpeed. We conducted pre-training tests for GPT3-7B and GPT3-175B on different scales of GPU nodes. As for the dataset, we used the C4 dataset, which consists of approximately 156B tokens. All experiments were conducted in the PyTorch 2.0 environment with CUDA 11.8.

We compared the TRANSOM system with three baseline methods as follows:

\textbf{Kubeflow\cite{bisong2019kubeflow}:} Kubeflow is an open-source platform developed by Google for machine learning and MLOps (Machine Learning Operations) on Kubernetes. For certain machine learning models and libraries, the Kubeflow Training Operator component provides support for Kubernetes custom resources. This component runs distributed or non-distributed TensorFlow, PyTorch, Apache MXNet, XGBoost, and MPI training jobs on Kubernetes. Several cloud platform companies, including Google, Microsoft, and Amazon, use Kubeflow for AI multi-node multi-GPU training.

\textbf{TORCH.SAVE \&\& TORCH.LOAD \cite{torch-save}:} torch.save and torch.load are functions provided by PyTorch for saving serialized objects to disk and deserializing files into memory, respectively. They can be used to save and load various objects, including models, tensors, and dictionaries. As the default checkpoint saving and loading method in PyTorch, they are used in various distributed parallel architectures like Megatron-LM, DeepSpeed, and Colossal-AI.

\textbf{Nebula \cite{nebula}:} Nebula is a fast, simple, diskless, and model-aware checkpointing tool in Azure Container for PyTorch (ACPT). Nebula offers a straightforward and high-speed checkpointing solution for distributed large-scale model training jobs using PyTorch. It is considered a state-of-the-art (STOA) solution.

\textbf{LOF (Local Outlier Factor) \cite{breunig2000lof}:} LOF anomaly detection algorithm is one of the most commonly used density-based anomaly detection algorithms. This method does not have strict assumptions about the distribution of data, making it robust and effective. It is widely used in various system anomaly analysis systems.

\subsection{Overall Performance}


In this subsection, we conducted a performance comparison of end-to-end training for GPT3-175B using both the Kubeflow and TRANSOM distributed training systems. Both experiments utilized 512 NVIDIA A800-80GB GPUs. The distributed training was achieved through a combination of tensor parallelism, pipeline parallelism, and data parallelism, with tensor parallelism restricted to dividing the GPUs within a single node (8 GPUs per node).

Furthermore, we measured the elapsed time from task submission to completing the training on the 300B dataset as the end-to-end training time for the training tasks. Comparative experiments were conducted using the distributed training system provided by Kubeflow and the TRANSOM distributed training system, which includes automatic anomaly detection and checkpoint asynchronous access mechanisms.

\begin{figure}[ht] 
    \centering 
    \captionsetup{justification=centering}
    \includegraphics[width=0.48\textwidth]{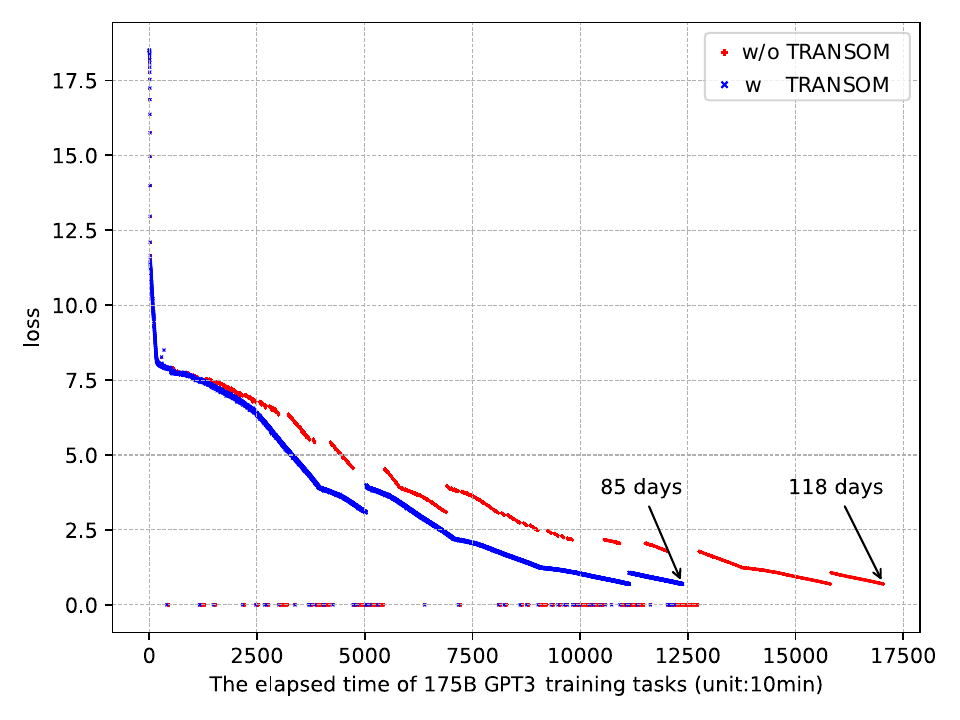}
    \caption{End-to-end Performance of LLMs training tasks. (where ``w/o" stands for without the TRANSOM system, and ``w/" stands for with the TRANSOM system)} 
    \label{fig:overall-perf} 
\end{figure}

The experimental results are shown in Figure \ref{fig:overall-perf}. We sampled the training task's lifecycle status by monitoring the loss values every 10 minutes. It can be observed that the large model training task without using the TRANSOM system took 118 days to complete training. In contrast, the LLMs task based on the TRANSOM system completed the training in only 85 days, resulting in a 28\% improvement in training efficiency. In the graph, there are many points with a loss value of 0, which can be attributed to two possibilities: one is the checkpoint saving process, and the other is task interruption.

The main reasons for the longer training time in tasks without the TRANSOM system are the additional time spent on checkpoint saving and the extended recovery time after interruptions. In some cases, the recovery time could be as long as 48-72 hours, especially during weekends and holidays. TRANSOM, on the other hand, utilizes its unique pipeline-based automatic state machine, real-time anomaly detection technology, and asynchronous checkpoint reading and writing methods. This allows TRANSOM to automatically filter out bad nodes and add new healthy nodes to the training without manual intervention, achieving an average task restart time of 10 minutes.

\subsection{Coverage of TEE}

To validate the effectiveness of TEE, we collected data from real online tasks over the course of a month. This dataset consisted of 13 instances of normal data and 11 instances of erroneous data. The normal data was used to train the TEE, and subsequently, the trained TEE was employed to make predictions on the erroneous data. It's important to note that we mentioned using three machine learning methods in \ref{anomaly-detection}. However, since DTW did not perform well in predicting erroneous tasks in the experiments, we implemented LOF and KNN matrix profile (referred to as Neighbor Profile, NProfile). In the following experiments, the predictive capabilities of LOF and NProfile are compared. If either method indicates an anomaly, we consider the task to be erroneous.

\begin{figure}[ht] 
    \centering 
    \captionsetup{justification=centering}
    \includegraphics[width=0.48\textwidth]{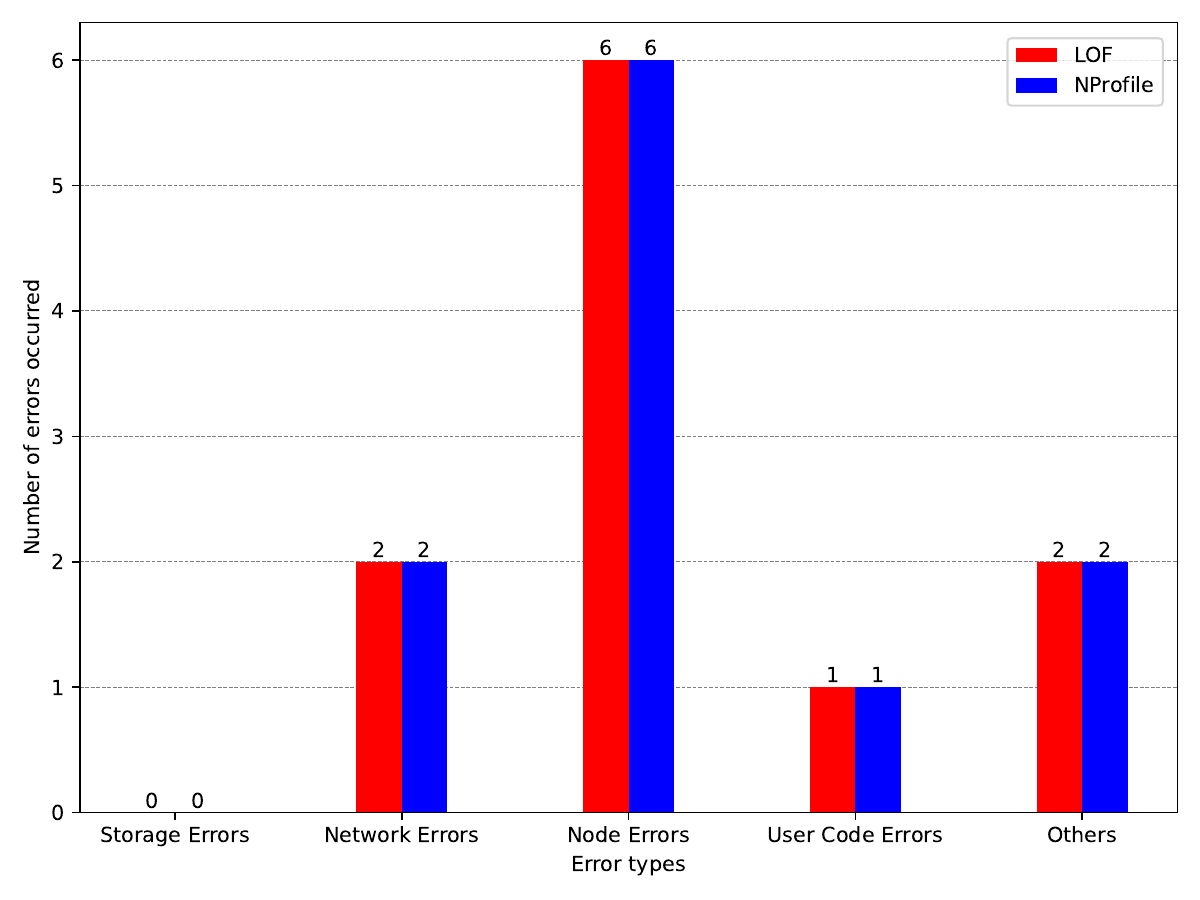}
    \caption{Number of errors with different kinds of types predicted by TEE in a month} 
    \label{fig:predict} 
\end{figure}

In figure \ref{fig:predict}, it's evident that both LOF and NProfile successfully predicted all 11 erroneous tasks. However, it's important to note that TEE is only applicable to LLMs tasks, specifically those with high GPU utilization and network traffic. It lacks the capability to accurately identify other tasks that do not meet these characteristics. For instance, if a user submits a ``sleep" task, it would also be considered an anomaly task.

\subsection{Benefits of TCE}

\begin{figure}[ht] 
    \centering 
    \captionsetup{justification=centering}
    \includegraphics[width=0.48\textwidth]{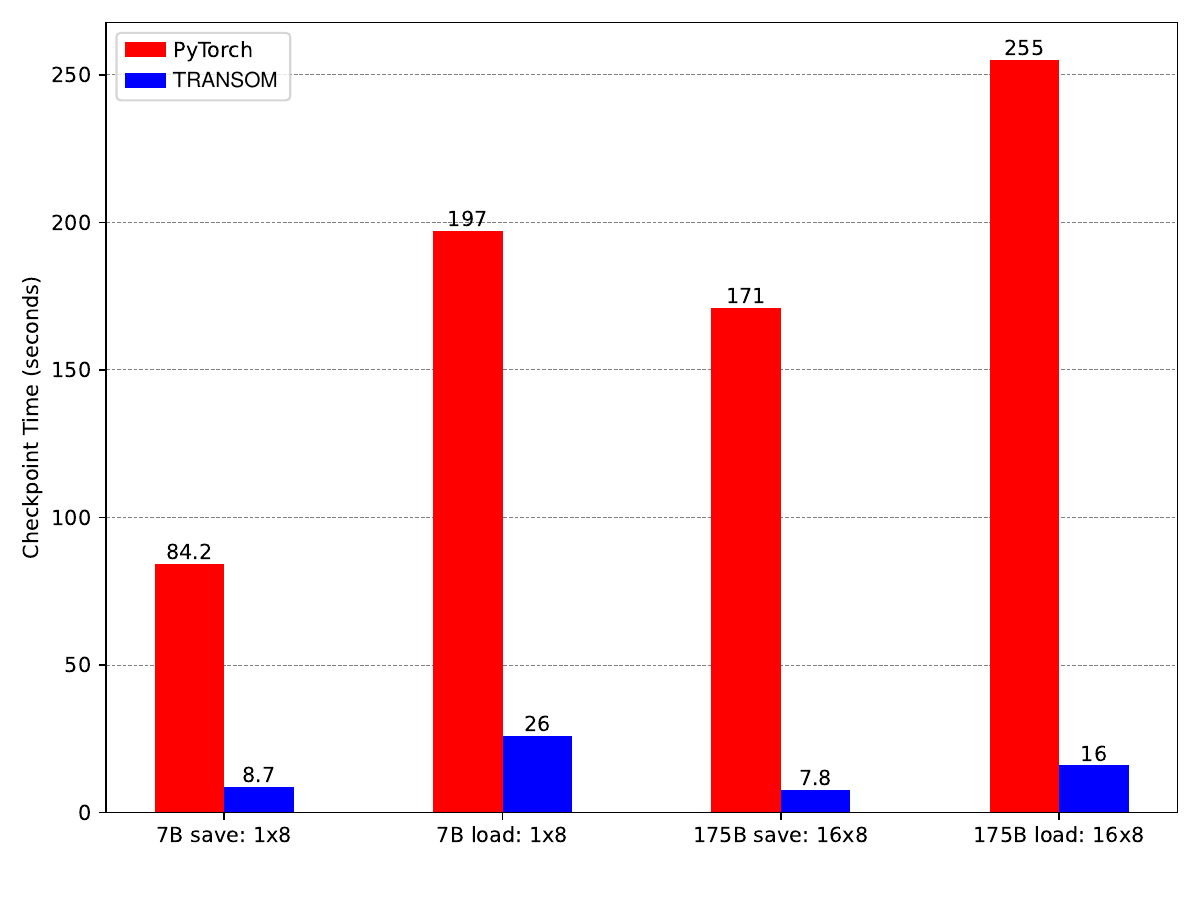}
    \caption{GPT3 Checkpoint Performance: PyTorch vs TRANSOM} 
    \label{fig:checkpoint-vs} 
\end{figure}

To demonstrate the performance of the checkpoint asynchronous reading and writing mechanism in the TRANSOM system, we conducted a comparison experiment for checkpoint read and write performance during the training processes of GPT3-7B and GPT3-175B on a single node with 8 A800-80GB GPUs and on 16 nodes. The experimental results are shown in Figure \ref{fig:checkpoint-vs}.

It can be observed that in the case of GPT3-7B, the save performance, based on the TRANSOM system's TCE checkpoint optimization mechanism, improved nearly 10 times, and the write performance increased by 7.5 times. For GPT3-175B, the read performance and write performance improved by 20 times and 16 times, respectively. This indicates a significant improvement in checkpoint I/O performance when using TRANSOM's TCE mechanism.

\begin{figure}[ht] 
    \centering 
    \captionsetup{justification=centering}
    \includegraphics[width=0.48\textwidth]{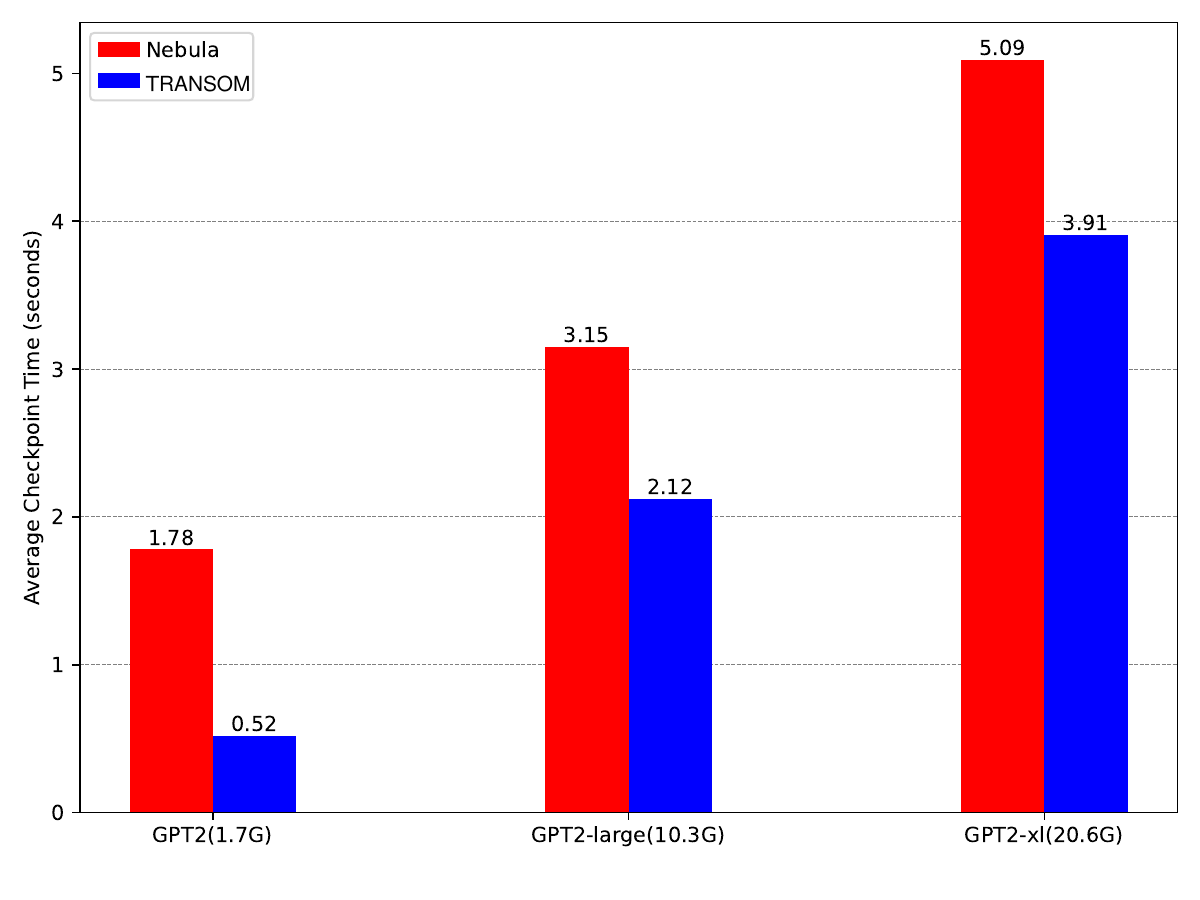}
    \caption{GPT2 Checkpoint Performance: Nebula vs TRANSOM} 
    \label{fig:ckpt-vs-nebula} 
\end{figure}

Additionally, to compare performance with Microsoft's state-of-the-art Nebula solution, we conducted tests on TRANSOM's average checkpoint saving performance for Hugging Face GPT2, GPT2-Large, and GPT2-XL models. Experimental results, as shown in Figure \ref{fig:ckpt-vs-nebula}, indicate that compared to Nebula, TRANSOM exhibits a performance improvement of 1.3 to 3.4 times when saving GPT2 models. It's worth noting that as the model size increases, TRANSOM's performance continues to show an upward trend.

\section{Conclusion}

We have introduced TRANSOM, a system designed for fault-tolerant initiation and execution of Large Language Models (LLMs) training tasks. The key idea is to provide automated preheating, anomaly detection, and fault recovery capabilities for large-scale pretraining processes using a task-level finite-state automaton mechanism. We utilize a multi-model decision prediction method for real-time anomaly detection during the training process and achieve efficient checkpoint read-write operations through an asynchronous redundancy mechanism.

In the design and testing of TRANSOM, we addressed significant technical challenges associated with frequent abnormal restarts, low fault diagnosis efficiency, and high checkpoint I/O overhead in large-scale LLMs training tasks. Leveraging multi-dimensional metric data during the training process and employing machine learning decision reasoning methods, we identified anomalies and faulty nodes in the training process. Through the pipeline fault-tolerance system, we interrupted the training tasks, terminated container processes on faulty nodes, and rescheduled containers to other healthy nodes for distributed training.

Furthermore, during the process of resuming training, the TRANSOM system adopts a mechanism where the original containers recover training by directly reading checkpoints from memory, while newly added containers recover training by pulling data from neighboring container nodes through RDMA channels. This approach significantly reduces the time required for task restart and recovery.

Through experiments conducted on both physical and simulated clusters, it has been demonstrated that the TRANSOM fault-tolerant system can reduce the execution time of GPT3-175B LLMs training tasks by 33 days. It achieves a recognition rate of over 100\% for common LLMs training anomalies and reduces the average read and write overhead for checkpoints from 255 seconds to 16 seconds per operation. This approach greatly enhances the execution efficiency of large-scale LLMs training tasks and reduces the cost of manual intervention and troubleshooting.


\bibliographystyle{IEEEtran}
\bibliography{refs}

\end{document}